\documentclass[twocolumn,showpacs,showkeys,prc,aps,amsmath,amssymb]{revtex4}
\usepackage{graphicx}
\usepackage{dcolumn}
\usepackage{bm}

\begin{document}

\title{Shell model analysis of competing contributions to the double-beta decay of $^{48}$Ca}

\author{Mihai Horoi}
 \email{mihai.horoi@cmich.edu}
\affiliation{Department of Physics, Central Michigan University, Mount Pleasant, Michigan 48859, USA}

\date{\today}

\begin{abstract}
\noindent
{\bf Background:} Neutrinoless double beta decay, if observed, would reveal physics beyond
the Standard Model (SM) of particle physics, namely it would prove that neutrinos are Majorana
fermions and that the lepton number is not conserved.\\
{\bf Purpose:} The analysis of the results of neutrinoless double beta decay observations
requires an accurate knowledge of several nuclear matrix elements (NME) for different mechanism
that may contribute to the decay. We provide a complete analysis of these NME for the decay
of the ground state (g.s.) of $^{48}$Ca to the g.s. $0^+_1$ and first excited $0^+_2$ state
of $^{48}$Ti.\\
{\bf Method:} For the analysis we used the nuclear shell model with effective two-body interactions
that were fine-tuned to describe the low-energy spectroscopy of $pf$-shell nuclei. We checked our 
model by calculating the two-neutrino transition probability to the g.s. of $^{48}$Ti.
We also make predictions for the transition to the first excited $0^+_2$ state of $^{48}$Ti.\\
{\bf Results:} We present results for all NME relevant for the neutrinoless transitions to 
the $0^+_1$ and $0^+_2$ states, and using the lower experimental limit for the g.s. to g.s. 
half-life we extract upper limits for the neutrino physics parameters.\\ 
{\bf Conclusions:} We provide accurate NME for the two-neutrino and neutrinoless double beta decay transitions in A=48 system, which can be further used to analyze the experimental results of 
double beta decay experiments when they become available.    
\end{abstract}

\maketitle

\section{Introduction}

Neutrinoless double beta ($0 \nu \beta \beta$) decay, which can only occur by violating the conservation of the total 
lepton number, if observed it will reveal physics beyond the Standard Model, and it will represent 
a major milestone in the study of the fundamental properties of neutrinos \cite{HAX84}-\cite{rmp-08}.
Indeed, its discovery would decide if neutrinos are their own antiparticles \cite{sv82}, and would provide a hint about the scale of their absolute masses.
That is why there are intensive investigations of this process, both theoretical and experimental.
Recent results from neutrino oscillation experiments have
demonstrated that neutrinos have mass and they can
mix \cite{kamland}-\cite{atmospheric}. However, the neutrino oscillations experiments
cannot be used to determine the neutrino mass hierarchy and the lowest neutrino mass.
Neutrinoless double beta decay is viewed as one of the best routes to decide
these unknowns.
 A key ingredient for extracting the absolute neutrino masses from
$0\nu\beta\beta$ decay experiments is a precise knowledge of the
nuclear matrix elements (NME) for this process.

There are potentially many mechanisms that could contribute to the neutrinoless double beta
decay process that will be briefly reviewed below. 
Several of these mechanisms do not provide contributions
to the decay rate that explicitly depend on the neutrino masses, but their effect would
vanish if the neutrinos are not massive Majorana particles \cite{sv82}. 
In all cases the half-lives depend on the nuclear matrix elements that need to be accurately
calculated using low-energy nuclear structure models. In particular, if the exchange of light 
left-handed neutrinos 
is proven to be the dominant mechanism, one could be able to use the experimental results and the
associated NME to extract the neutrino mass hierarchy  and the lowest neutrino mass \cite{rmp-08}.
The two-neutrino double beta ($2 \nu \beta \beta$) decay is an associate process that is allowed 
by the Standard Model, and it was observed in about ten isotopes. Therefore, a good but
not sufficient test of nuclear structure models would be a reliable description of the 
$2 \nu \beta \beta$ half-lives.
 
Since most of the $\beta\beta$ decay emitters are open shell nuclei,
many calculations of the NME have been performed within the pnQRPA approach and its 
extensions \cite{VOG86}-\cite{sim-97}. However, the pnQRPA calculations of the
more common two-neutrino double beta decay half-lives, which were measured for about
10 cases \cite{barabash10},
are very sensitive to the variation of the so called $g_{pp}$ parameter (the strength 
of the particle-particle interactions in the $1^+$ channel) \cite{VOG86}-\cite{SUH88}, and this 
drawback still persists in spite of various improvements brought by its 
extensions \cite{SMK90}-\cite{BKZ00}, including higher-order QRPA approaches \cite{RFS91}-\cite{sim-97}. 
The outcome of these attempts was that the calculations became more stable against $g_{pp}$ variation,
but 
at present there are still large differences between the values of the NME calculated with different 
QRPA-based methods, which do not yet provide a reliable determination of the 
two-neutrino double beta decay half-life.
Therefore, although the QRPA methods do not seem to be suited to predict the $2\nu\beta\beta$ 
decay half-lives, one
can use the measured $2\nu\beta\beta$ decay half-lives to calibrate the $g_{pp}$ parameters, which are further used
to calculate the $0\nu\beta\beta$ decay NME \cite{rodin07}. 
Other methods that were recently used to provide NME for most $0\nu\beta\beta$ decay
cases of interest are the Interacting Boson Model (IBM-2) \cite{iba-09,iba-12}, 
the Projected Hatree-Fock Bogoliubov (PHFB) \cite{phfb}, and
the Generator Coordinate Method (GCM) \cite{gcm}. 

Recent progress in computer power,  
numerical algorithms, and improved nucleon-nucleon effective interactions, made possible
large scale shell model calculations (LSSM) of the $2\nu\beta\beta$ and $0\nu\beta\beta$ decay
NME \cite{plb-ca48}-\cite{retamosa-95}.
The main advantage of the large scale shell model  calculations is that they seem to be 
less dependent on the effective interaction used, as far as these interactions are consistent
with the general spectroscopy of the nuclei involved in the decay. Their main drawback is the 
limitation imposed by the exploding shell model dimensions on the size of the valence
spaces that can be used.
The most important success of the large scale shell model calculations 
was the correct prediction of the $2\nu\beta\beta$ decay 
half-life for $^{48}$Ca \cite{plb-ca48,exp-ca48}. 
In addition, these calculations did not have to adjust any additional parameter, i.e.
given the effective interaction and the Gamow-Teller (GT) quenching factor extracted 
from the overall spectroscopy in the mass-region (including beta decay probabilities 
and charge-exchange strength functions), one can reliably predict the $2\nu\beta\beta$ decay half-life of $^{48}$Ca.

Clearly, there is a need to further check and refine these calculations, and to provide
more details on the analysis of the NME that could be validated by experiments.
We have recently revisited \cite{HSB07} the $2 \nu \beta \beta$ decay of
$^{48}$Ca using two recently proposed effective interactions for this mass region,
GXPF1 and GXPF1A, calculating the NME and half-lives for the
transition of the $^{48}$Ca g.s. to the g.s. and the first excited $2^{+}$ state of
$^{48}$Ti. 

In this paper we add to the analysis the $2\nu\beta\beta$ transition to the
first excited $0^+_2$ state of $^{48}$Ti. We also extend our analysis \cite{mh10-prc2}
of the $0\nu\beta\beta$
decay of $^{48}$Ca by
providing the NME associated with the most important 
 $0\nu\beta\beta$ mechanisms for
transitions to the g.s. $0^+_1$ and first excited $0^+_2$ state
of $^{48}$Ti.
Future experiments on double beta decay of $^{48}$Ca (CANDLES \cite{candles-06} and
CARVEL \cite{carvel-05}) may reach the required sensitivity of measuring such transitions, and
our results could be also useful for planning these experiments.

\section{Two-neutrino double beta decay}

LSSM calculations of $2 \nu \beta \beta$  decay NME can now be 
carried out rather accurately for many nuclei \cite{caur-136xe}. In the case of $^{48}$Ca,
 Ref. \cite{plb-ca48} reported for the first time
a full $pf$-shell calculation of the NME for the $2\nu\beta\beta$ decay mode, 
for both transitions to the g.s. and to the $2^+_1$ excited state of $^{48}$Ti. 
As an effective interaction it was   
used the Kuo-Brown G-matrix \cite{[KB68]} with minimal 
monopole modifications, KB3 \cite{[PZ81]}. 
 In Ref. \cite{HSB07} we use the recently proposed GXPF1A two-body effective interaction, 
which has been successfully 
tested for the $pf$ shell \cite{[gx1ap]}-\cite{[ni56]}, to perform $2\nu\beta\beta$ decay 
calculations for $^{48}$Ca. 
Our goal was to obtain the values of  NME for this decay mode, for both transitions to the 
g.s. and to the $2^+_1$ state of $^{48}$Ti, with increased degree of confidence, 
which would allow us to consider similar calculations for the $0\nu\beta\beta$ decay mode of 
this nucleus \cite{retamosa-95}.  
The $2\nu\beta\beta$ transitions to excited states have longer half-lives, as compared
with the transitions to the g.s., due to the reduced values of the
corresponding phase space factors, but they were measured in some cases, such as
 $^{100}$Mo \cite{mo05}.


For the $2\nu\beta\beta$ decay mode the relevant NME are of Gamow-Teller type,  
and has the following expression for  decays to states in the grand-daughter that
have the angular momentum $J=0,2$ \cite{HAX84}-\cite{srepv02},

\begin{equation}
M^{2\nu}_{GT}(J^+) = \frac{1}{\sqrt{J+1}}
\sum_k \frac{\langle J^+_f\vert\vert \sigma\tau^-\vert\vert 1^+_{k}
\rangle 
\langle 1^+_{k}\vert\vert \sigma\tau^-
\vert\vert 0^+_i\rangle}{(E_k + E_J)^{J+1}} .
\label{eq1}
\end{equation}

\noindent 
Here $E_k$ is the excitation energy of the $1^+_k$ state of intermediate odd-odd nucleus, 
and   
$E_J = \frac{1}{2}Q_{\beta\beta}(J^+) + \Delta M$.
 $Q_{\beta\beta}(J^+)$ is the Q-value 
corresponding to the $\beta\beta$ 
decay to the final $J^+_f$  state of the grand-daughter nucleus, 
 and $\Delta M$ is the  mass difference between the parent and the intermediate
nucleus $^{48}$Sc.  
The most common case is the decay to the $0^+_1$ g.s. of the grand-daughter, but
decays to the first excited $0^+_2$ and $2^+_1$ states are also investigated.

The $2\nu\beta\beta$ decay  half-life expression is given by

\begin{equation}
\left[T^{2\nu,J}_{1/2}\right]^{-1} = 
G^{2\nu}_J\vert M^{2\nu}_{GT}(J)\vert^2 
\label{hlive}
\end{equation}

\noindent
where $G^{2\nu}_J$ are $2 \nu \beta \beta$ phase space factors. Specific values of $G^{2\nu}_J$ for
different $2\nu\beta\beta$ decay cases can be found in different reviews, such as Ref. \cite{SC98}.
For a recent analysis of $G^{2\nu}_J$ see Ref. \cite{kipf12}. 
In Ref. \cite{HSB07} we explicitly analyzed the dependence of the double-Gamow-Teller
sum entering the NME Eq. (\ref{eq1}) vs the excitation energy of the $1^+$ states in the 
intermediate nucleus $^{48}$Sc. This sum was recently investigated 
experimentally \cite{prl-0709}, and it was shown that indeed, the incoherent sum 
(using only absolute values of the Gamow-Teller matrix elements) would
provide an incorrect NME, thus validating our prediction.
We have also corrected by several orders of magnitude the probability of
transition of the g.s. of $^{48}$Ca to the first excited $2^{+}$ state of
$^{48}$Ti reported in Ref. \cite{plb-ca48}.

\begin{table}[tbe]
	\caption{Matrix elements and half-lives for $2\nu$ decay 
calculated using GXPF1A interaction and two quenching
factors. Matrix elements are in MeV$^{-1}$ for transitions to $0^+$ states and in MeV$^{-3}$ for transitions to 
$2^+$ states. } 
	\begin{center}
        \begin{tabular}{|c|c|c|c|c|}
            \hline
 & \multicolumn{2}{c|}{} & \multicolumn{2}{c|}{} \\
            & \multicolumn{2}{c|}{$qf=0.77$} & \multicolumn{2}{c|}{$qf=0.74$} \\
	 \cline{2-5}
 & & & & \\
   $J^{\pi}_n$ & $M^{2\nu}$ & $T^{2\nu}_{1/2}$ (y) & $M^{2\nu}$ & $T^{2\nu}_{1/2}$ (y) \\
	\hline
 & & & & \\
          $0^+_1$ &  0.054  & $3.3\times 10^{19}$ & 0.050 & $3.9\times 10^{19}$\\
& & & & \\
      $2^+_1$ &  0.012  & $8.5\times 10^{23}$ & 0.010 & $1.0\times 10^{24}$\\
& & & & \\
      $0^+_2$ &  0.050  & $1.6\times 10^{24}$ & 0.043 & $1.9\times 10^{24}$\\
            \hline
        \end{tabular}
	\end{center}
\end{table}

In Ref. \cite{HSB07} we fully diagonalized 250 $1^+$ states in the intermediate nucleus to calculate
the $2 \nu \beta \beta$ decay NME for $^{48}$Ca. This procedure can be used for somewhat heavier nuclei
using the J-scheme shell model code NuShellX \cite{nushellx}, 
but for cases with large dimension one needs an alternative
method. The pioneering work on $^{48}$Ca \cite{plb-ca48} used a strength-function approach that converges
after a small number of Lanczos iterations, but it requires large scale shell model diagonalizations when one 
wants to check the convergence. Ref. \cite{ehv92} proposed an alternative method, which converges very 
quickly, but it did not provide a complete recipes for all its ingredients, 
and it was never used in practical calculations. 
Recently \cite{mh-cssp10}, we proposed a simple numerical scheme to calculate all coefficients of the expansion 
proposed in Ref. \cite{ehv92}.
Following Ref. \cite{ehv92}, we choose as a starting Lanczos vector, $L^{\pm}_1$, 
either the initial or final state in the 
decay (only $0^+$ to $0^+$ transitions are considered), to which we apply the Gamow-Teller operator.
This approach is very efficient for large model spaces, 
as for example the $jj55$ space (consisting of the
$0g_{7/2}$, $1d, 2s$, and $h_{11/2}$ orbits), which for the $^{128}$Te decay leads to m-scheme dimensions
of the order of 10 billions necessary to calculate the g.s. of $^{128}$Xe.
In the calculation of $^{48}$Ca decay we use the standard quenching factor, 
$qf=0.77$, for the Gamow-Teller operator $\sigma \tau$. We checked the result reported in Ref. \cite{HSB07} 
using this alternative method and we found the same result. The novel result report here for the
first time is for the transition to the first excited $0^+$ state in $^{48}$Ti at 2.997 MeV. The 
matrix element when using GXPF1A interaction 
is 0.050, very close to that for the transition to the g.s. Using the phase
space factor $G^{2\nu}_{0^+_2} = 2.43\times 10^{-22} \ MeV^{-1}$ from Ref. \cite{SC98} 
(a new set of phase space
factors were recently proposed \cite{kipf12}, but for $2\nu\beta\beta$ 
decays they differ only by 4\% from 
those of Ref. \cite{SC98})
 we found that the half-life for this transition is
$1.6 \times 10^{24}$ y. We recall here that our results reported in \cite{HSB07} 
for the half-lives of the transitions to g.s. and to the first $2^+$ excited state are 
 $3.3 \times 10^{19}$ y and $8.5 \times 10^{23}$ y, respectively. One can see that the transition 
to the first excited $0^+_2$ state at 2.997 MeV is predicted to compete with the 
transition to the first excited $2^+_1$ state at 0.994 MeV. 

The half-life for the transition
to the g.s. $0^+_{1}$ was measured by several groups with increased precision (see e.g. \cite{barabash10}).
The most recent result from NEMO-3 (see \cite{barabash10} and references therein) 
is $T^{2\nu}_{1/2}=4.4^{+0.5}_{-0.4}(stat.)\pm 0.4(syst.)$.
Our GXPF1A result is marginally out of the recently reduced error bars. 
However, a recent publication \cite{qf12} found a quenching factor of 0.74 for the
$pf$-shell nuclei using GXPF1A interaction. The same quenching factor was proposed some
time ago \cite{qf96} using a different effective interaction.  
Using the smaller quenching factor of 0.74 brings the 
calculated half-life within the experimental limits. A comparison of the matrix elements and the
associated half-lives for the two quenching factors used here is given in Table I. 
Potential observation of the $2 \nu \beta \beta$
transitions to the excited states of $^{48}$Ti could shed some light on the variation of 
the quenching factor for the Gamow-Teller operator in this nucleus. One should also mention
that the excitation energy of the $0^+_2$ state in $^{48}$Ti calculated with GXPF1A interaction is
about 1 MeV higher than the experimental value, while it is about right for $^{48}$Ca.
Other available effective interactions do no provide a better description of this state.
This result may raise concerns about the validity of the nuclear structure description of this state
within the $pf$-shell. An experimental observation of the $2 \nu \beta \beta$ transition to this
state could be used to validate (or not) our result.


\section{Neutrinoless double beta decay}

The $0 \nu \beta \beta$ decay, $(Z,A) \rightarrow (Z+2,A)+2e^-$, requires the neutrino to be a 
massive Majorana fermion, i.e. it is identical to the antineutrino \cite{sv82}.  We already know from
the neutrino oscillation experiments that some of the neutrinos participating 
in the weak interaction have mass, and that
the mass eigenstates 
are mixed by the PNMS matrix $U_{lk}$, where $l$ is the lepton flavor and $k$ is the mass eigenstate number (see e.g. Ref. \cite{ves12}). However, 
the neutrino oscillations experiments cannot decide the mass hierarchy, the mass of the lightest neutrino,
and some of the CP non-conserving phases of the PNMS matrix (assuming that neutrinos are
Majorana particles).

Considering only contributions from the exchange of light, left-handed(chirality),  Majorana neutrinos \cite{rmp-08}, the $0\nu\beta\beta$ decay half-live is given by

\begin{equation}
\left[ T^{0\nu}_{1/2} \right]^{-1}=G^{0\nu}\left| M^{0\nu}_{\nu}\right|^2 
\left( \frac{ \mid \left<  m_{\beta \beta} \right> \mid}{m_e} \right)^2 \ .
\label{hll}
\end{equation}

\noindent
Here, $G^{0\nu}$ is the phase space factor, which depends on the $0\nu\beta\beta$ decay energy,
$Q_{\beta\beta}$, the charge of the decaying nucleus Z, and the nuclear radius \cite{SC98,kipf12}. 
The effective neutrino mass, $\left< m_{\beta \beta} \right>$, is related to the neutrino mass eigenstates, 
$m_k$, via the left-handed lepton mixing matrix, $U_{e k}$,

\begin{equation}
\left< m_{\beta \beta} \right>/m_e \equiv \eta_{\nu L} =  \sum_{k=light} m_{k} U^{2}_{e k}  \ /m_e.
\label{etanl}
\end{equation}

\noindent
$m_e$ is the electron mass. The NME, $M^{0\nu}_{\nu}$, is given by
\begin{equation}
 M^{0 \nu}_{\nu}=M^{0 \nu}_{GT}-\left( \frac{g_V}{g_A} \right)^2  M^{0 \nu}_F- M^{0 \nu}_T\ ,
\label{sme}
\end{equation}

\noindent
where $M^{0 \nu}_{GT}$, $M^{0 \nu}_F$ and $M^{0 \nu}_T$ are the Gamow-Teller (GT), 
Fermi (F) and tensor (T) matrix elements, 
respectively. Using closure approximation these matrix elements are defined as follows:

\begin{widetext}
\begin{eqnarray}
\nonumber 
M_\alpha^{0\nu} & = & \left< 0^+_f \mid \sum_{m,n} \tau_{-m} \tau_{-n} O^\alpha_{mn} \mid 0^+_i \right> \\
 & = &
\sum_{j_p j_{p^\prime} j_n j_{n^\prime} J_\pi} TBTD 
\left( j_p j_{p^\prime} , j_n j_{n^\prime} ; J^\pi \right) 
\left< j_p j_{p^\prime}; J^\pi T \mid \tau_{-1} \tau_{-2}O^\alpha_{12} 
\mid j_n j_{n^\prime} ; J^\pi T\right>_a ,
\label{mbme}
\end{eqnarray}
\end{widetext}

\noindent
where $O^\alpha_{mn}$ are $0\nu\beta\beta$ transition operators, $\alpha=(GT,\ F,\ T)$, $\mid 0^+_i >$
is the g.s. of the parent nucleus, 
and  $\mid 0^+_f >$ is the final $0^+$ state of the
grand daughter nucleus.
The two-body transition densities (TBTD) can be obtained from LSSM calculations \cite{mh10-prc2}.
Expressions for the anti-symmetrized two-body matrix elements (TBME) 
$ \left< j_p j_{p^\prime}; J^\pi T \mid \tau_{-1} \tau_{-2}O^\alpha_{12} \mid j_n j_{n^\prime} ; J^\pi T\right>_a $
can be found elsewhere, e.g. Refs. \cite{nsh12,mh10-prc2}. 
Assuming that one can unambiguously measures a $0 \nu \beta \beta$ half-life, and one can reliably
calculate the NME for that nucleus, one could use Eqs. (\ref{hll}) and (\ref{etanl}) to extract 
information about the lightest neutrino mass and the neutrino mass hierarchy \cite{ves12}.
In addition, one could consider the contribution from the right-handed currents to the effective Hamiltonian, 
which can mix light and heavy neutrinos of both chiralities (L/R)

\begin{eqnarray}
\nonumber \nu_{eL}=\sum_{k=light} U_{ek} \nu_{kL}+\sum_{k=heavy} U_{ek} N_{kL}\\
\nu_{eR}=\sum_{k=light} V_{ek} \nu_{kR}+\sum_{k=heavy} V_{ek} N_{kR}\ ,
\end{eqnarray}

\noindent
where $N_k$ are the heavy neutrinos that are predicted by several see-saw mechanisms 
for neutrino masses \cite{ves12}. $U_{lk}$ and $V_{lk}$ are the left and right-handed components
of the unitary matrix that diagonalizes the neutrino mass matrix \cite{valle12}.
One should also mention that there are several other
mechanisms that could contribute to the $0 \nu \beta \beta$ decay, such as the exchange of supersymmetric
(SUSY) particles (e.g. gluino and squark exchange \cite{prd83}), etc, 
whose effects are not directly related to the neutrino masses, but indirectly via the
Schechter-Valle theorem \cite{sv82}. Assuming that the masses of the light neutrinos are smaller than 
1 MeV and the masses of the heavy neutrinos, $M_k$, are larger than 1 GeV, the particle physics 
and nuclear structure parts get separated, and the inverse half-life can be 
written as

\begin{widetext}
\begin{eqnarray}
\nonumber
\left[ T^{0\nu}_{1/2} \right]^{-1}  & =  G^{0\nu} \left| \eta_{\nu L} M^{0 \nu}_{\nu} \right. & 
+ <\lambda> \tilde{X}_{\lambda} 
+ <\eta> \tilde{X}_{\eta}  +
\left(\eta_{N L} + \eta_{N R}\right) M^{0 \nu}_N \\
 & & + \ \  \left. \eta_{\lambda'} M^{0 \nu}_{\lambda'} 
 + \eta_{\tilde{q}} M^{0 \nu}_{\tilde{q}}
 + \eta_{KK} M^{0 \nu}_{KK} \right|^2,
\label{t0gen}
\end{eqnarray}
\end{widetext}

\noindent
where $\eta_{\nu L}$ was defined in Eq. (\ref{etanl}), and

\begin{eqnarray}
\nonumber
\eta_{N L} & = & \sum_{k=heavy} U_{ek}^2 \frac{m_p}{M_k}, \\ 
\nonumber
\eta_{N R} & \approx & 
\left(\frac{M_{W_L}}{M_{W_R}}\right)^4 \sum_{k=heavy} V_{ek}^2 \frac{m_p}{M_k}, \\
\nonumber
  <\lambda> & = & \epsilon \sum_{k=light} U_{ek}V_{ek},\\  
<\eta> & = &\left( \frac{M_{WL}}{M_{WR}} \right)^2 \sum_{k=light} U_{ek}V_{ek}\ .
\label{etas}
\end{eqnarray}

\noindent
Here $\epsilon$ is the mixing parameter for the right heavy boson $W_R$ and the standard left-handed heavy boson $W_L$, $W_R \approx \epsilon W_1 + W_2$, $M_{WR}$ and $M_{WL}$ are their respective masses, 
and $m_p$ is the proton mass. The $\eta_{\lambda'}$ and $\eta_{\tilde{q}}$ are the R-parity violation 
contributions in supersymmetric (SUSY) Grand Unified Theories (GUT) 
related to the long range gluino exchange
and squark-neutrino mechanism, respectively \cite{ves12}. Finally, the $\eta_{KK}$ term is due to possible
Kaluza-Klein (KK) neutrino exchange in an extra-dimensional model \cite{KK03}.
The set of nuclear matrix elements $M^{0 \nu}_{\nu}$, $\tilde{X}_{\lambda}$, 
$\tilde{X}_{\eta}$, $M^{0 \nu}_N$, $M^{0 \nu}_{\lambda'}$, and $M^{0 \nu}_{\tilde{q}}$
 are discussed in many reviews, e.g. Ref. \cite{ves12}. The $M^{0 \nu}_{KK}$ analysis can be found 
in Ref. \cite{KK03}. In particular, using the factorization ansatz \cite{KK03} one gets

\begin{eqnarray}
\nonumber
\eta_{KK} M^{0 \nu}_{KK} & = & \frac{<m>_{SA}}{m_e}  M^{0 \nu}_{\nu} + m_p <m^{-1}> M^{0 \nu}_N \\
 & \equiv & \eta_{lKK} M^{0 \nu}_{\nu} + \eta_{hKK} M^{0 \nu}_N ,
\label{mkk}
\end{eqnarray}

\noindent
where $<m>_{SA}$ and $<m^{-1}>$ KK masses depend on the brane shift and bulk radius parameters, and
 are given in Table II of \cite{KK03}. One can see that the mass parameters $<m>_{SA}/m_e$ and 
$m_p <m^{-1}>$ has the effect of modifying $\eta_{\nu L}$ and $\eta_{N R}$ respectively.
$\mid m_p <m^{-1}> \mid < 10^{-8}$ and it could in principle compete with $\eta_{N R}$. 
$\mid <m>_{SA}/m_e \mid$ varies significantly with the model parameters and it could
also compete with $\eta_{\nu L}$.  One needs to go beyond the factorization ansatz, and use 
information from several nuclei \cite{KK07} 
to discern any significant contribution from the KK mechanism.

Constraints from colliders experiments suggest that terms proportional with the mixing angles,
$\epsilon$, $U_{ek(heavy)}$, and $V_{ek(light)}$ are very small \cite{valle12}. 
The present limits are $\mid <\lambda>\mid < 10^{-8}$ and $\mid <\eta>\mid <  10^{-9}$, but they
are expected to be smaller. In addition,
the contributions from $\tilde{X}_{\lambda}$ and $\tilde{X}_{\eta}$ terms in Eq. (\ref{t0gen})
would produce angular and energy distribution of the outgoing
electrons different than that coming from all other terms \cite{DKNT83}, and these signals 
are under investigation at SuperNEMO \cite{snemo10}. 
Here we assume that these contributions are small and can be 
neglected. In addition, if $<\lambda>$  is small, Eq. (\ref{etas}) suggests that 
$\eta_{N L}$ is small.
Information from colliders also puts some limits on 
$\left(M_{W_R},\ M_N\right) \sim \left(2.5 GeV,\ 1.4 GeV\right)$, and these limits will be
refined at LHC in the coming years.
Based on this information and the present limit on the $0 \nu \beta \beta$ decay of $^{76}$Ge one
can estimate that $\mid \eta_{\nu L}\mid < 10^{-6}$, and $\mid \eta_{N R}\mid < 10^{-8}$.
Then, the half-life can be written as


\begin{widetext}
\begin{eqnarray}
\left[ T^{0\nu}_{1/2} \right]^{-1}   =  G^{0\nu}\left| \tilde{\eta}_{\nu L} M^{0 \nu}_{\nu} 
 +  \tilde{\eta}_{N} M^{0 \nu}_N 
 +  \eta_{\lambda'} M^{0 \nu}_{\lambda'} 
 +  \eta_{\tilde{q}} M^{0 \nu}_{\tilde{q}} \right|^2,
\label{t0red}
\end{eqnarray}
\end{widetext}

\noindent
where we adjusted $\eta_{\nu L}$ and $\eta_{N R}$ for potential KK contributions,
$\tilde{\eta}_{\nu L}=\eta_{\nu L}+\eta_{lKK}$ and
$\tilde{\eta}_{N}=\eta_{N R}+\eta_{hKK}$.

If one neglects the SUSY and KK contributions until a hint of their 
existence is provided by colliders experiments or future results of $0\nu\beta\beta$ decay 
experiments show that these contributions are necessary \cite{KK07}, then

\begin{equation}
\left[ T^{0\nu}_{1/2} \right]^{-1} = G^{0\nu} \left(  \left| M^{0 \nu}_{\nu}
 \right|^2 \left| \eta_{\nu L}\right|^2 
+ \left| M^{0 \nu}_N \right|^2 \left|  \eta_{N R} \right|^2 \right) \ ,
\label{t0sred}
\end{equation}

\noindent
where we used the fact that the interference between the left-handed terms and the 
right-handed terms is negligible \cite{ves12}.

\begin{table}[tbe]
	\caption{Matrix elements  for $0\nu$ decay using GXPF1A interaction 
and two SRC models \cite{srcs09}, CD-Bonn (SRC1) and Argonne (SRC2). For comparison, the
$(a)$ values are taken 
from Ref. \cite{iba-12}, and the $(b)$ value is taken from Ref. \cite{wks99} 
for $g_{pp}=1$ and no SRC.} 
	\begin{center}
        \begin{tabular}{|c|c|c|c|c|c|}
            \hline
 \multicolumn{2}{|c|}{} & & & & \\
  \multicolumn{2}{|c|}{} & $M^{0 \nu}_{\nu}$ & $M^{0 \nu}_N$ & $M^{0 \nu}_{\lambda'}$
& $M^{0 \nu}_{\tilde{q}}$  \\
\hline
 & & & & & \\
$0^+_1$ & SRC1 &  0.90 & 75.5 & 618 & 86.7 \\
& & & & & \\
 & SRC2 & 0.82 & 52.9 & 453 & 81.8 \\
 \cline{2-6}
 & & & & & \\
  &  others & 2.3$^{(a)}$ & 46.3$^{(a)}$ & 392$^{(b)}$ & \\      

\hline      
 & & & & & \\
$0^+_2$ & SRC1 &  0.80 & 57.2 & 486 & 84.2 \\
& & & & & \\
 & SRC2 & 0.75 & 40.6 & 357 & 80.6 \\
            \hline
        \end{tabular}
	\end{center}
\end{table}

 The structure of the $M^{0 \nu}_N$ is the same as that described in Eqs. (5)-(8), with slightly
different neutrino potentials $H_{\alpha}(r)$ (see e.g. page 68 of Ref. \cite{ves12}).  
A detailed description of the matrix elements of $O^\alpha_{12}$ for the $jj$-coupling 
scheme consistent with the
conventions used by modern shell model effective interactions is 
given in Ref. \cite{mh10-prc2}.
One should also mention that our method \cite{mh10-prc2} of calculating the TBTD. Eq. (\ref{mbme}), is
different from that used in other shell model calculations \cite{retamosa-95}.
We included in the calculations the recently proposed higher order terms
of the nucleon currents, three old and recent parametrization of the short-range correlations (SRC) effects,
finite size (FS) effects, intermediate states energy effects, and we treated careful few other 
parameters entering the into the calculations.
We found very small variation of the NME with the average energy of the intermediate states, 
and FS cutoff parameters, and moderate variation vs the effective interaction and SRC
parametrization. 
We could also show that if the ground state wave functions of the initial and final 
nucleus can be accurately described using only the valence space orbitals, 
the contribution from the core orbitals can be neglected. This situation is different from that 
of the nuclear parity-nonconservation matrix elements \cite{pnc-prl-hb}, for which the "mean-field"
type contribution from the core orbitals could be significant \cite{ax-pncrev}. 
Another important result that clearly transpires from our formalism is that
in the closure approximation the neutrinoless transition to the first excited 
$2^+$ state is zero. This result is due to the rotational invariance of the TBME
entering Eq. (\ref{mbme}) (see also Appendix of Ref. \cite{mh10-prc2}).
The structure of the R-parity breaking SUSY mechanisms  NME is similar to that of light
and heavy neutrino exchange mechanisms,
but with no $\alpha=F$ component \cite{prd83}. 
The neutrino potentials used here for the $M^{0 \nu}_N$, 
and those used for the most significant contributions to 
$M^{0 \nu}_{\lambda'}$ and $M^{0 \nu}_{\tilde{q}}$ NME are given in
Ref. \cite{ves12}, but for completeness they are reviewed in the Appendix with the 
specific parameters included in these calculations.

\begin{table}[tbe]
	\caption{Single mechanism upper limits for neutrino physics 
parameters $\eta_j$ extracted
from the lower limit of the half-life for the transition to the ground state 
of $^{48}$Ti \cite{ves12} and using the matrix elements from Table II.} 
	\begin{center}
        \begin{tabular}{|c|c|c|c|c|c|}
            \hline
 \multicolumn{2}{|c|}{} & & & & \\
  \multicolumn{2}{|c|}{} & $\left| \tilde{\eta}_{\nu L}\right|\times10^5$ 
& $\left| \tilde{\eta}_{N}\right|\times10^7$ 
& $\left| \eta_{\lambda'}\right|\times10^8$ & $\left| \eta_{\tilde{q}}\right|\times10^7$  \\
\hline
 & & & & & \\
$0^+_1$ & SRC1 & 3.79 &	4.52 & 5.52 & 3.93  \\
& & & & & \\
 & SRC2 & 4.16 & 6.45 & 7.53 & 4.17 \\
            \hline
        \end{tabular}
	\end{center}
\end{table}

The results for all NME entering Eq. (\ref{t0red}) for the 
transition to the $0^+_1$ g.s. and first excited $0^+_2$ state of $^{48}$Ti
are presented in Table II.
Comparison with results of other models, when available, are also included.
For the light neutrino exchange matrix element we choose to compare with the
IBA-2 results, which is very different from ours. Other shell model analyses of this
particular NME  gives similar results for both transitions to $0^+_1$ and $0^+_2$ states
\cite{menendez-npa818,retamosa-95}.
        To our knowledge, with the exception of the light
neutrino exchange NME, no other results of
shell model calculations for these matrix elements were reported so far
(with the possible exception of Ref. \cite{menendez-h} where the NME
as a function of neutrino mass is reported and it could potentially be used to extract
the corresponding $M^{0 \nu}_N$). 
Based on these calculations and using the experimental 
lower limit of the half-life,
one can extract the "single-mechanism dominance" upper limits for $\mid \eta_j\mid$,
where $j=(\nu L),\ N, \ \lambda',\ \tilde{q}$. 
At present there is available  only the  lower
limit of the half-life for the transition to the g.s. of $^{48}$Ti,  
$1.4\times10^{22}$ y \cite{ves12}.
Using the phase-space factor from Ref. \cite{kipf12}, $G^{0 \nu}=61.4\times10^{-15}$
y$^{-1}$ (for $g_A=1.254$ and $R=1.2 A^{1/3}$ fm), 
we obtained the upper limits for $\mid \eta_j\mid$ shown in Table III.
Alternatively, assuming
that two or more mechanisms are contributing to the half-life in 
Eq. (\ref{t0red}) compete, 
one could use the experimental data from several isotopes to assess the contribution 
of each mechanism \cite{prd83}. Clearly, this scenario requires as many as possible
accurate half-lives and associated NMEs. For example, in the likely scenario that no
more than two mechanisms are competing and they are the light and heavy neutrino exchange,
then Eq. (\ref{t0sred}) can be used to analyze the data.
If the exchange of light neutrino will be determined as the dominant mechanism, then our
results could possible be used to decide the light neutrino mass hierarchy and the 
lowest neutrino mass \cite{ves12}.


\section{Conclusions and outlook}

In conclusion, we  analyzed the $2 \nu \beta \beta$ and several mechanisms that could compete
to the $0 \nu \beta \beta$ decays of $^{48}$Ca
 using shell model techniques. We described very efficient techniques 
to  calculate accurate
$2 \nu \beta \beta$ NME for  cases that involve large shell model dimensions. 
These techniques were tested for the case of $^{48}$Ca, and we provided NME and half-lives
for $2 \nu \beta \beta$ transitions to the g.s. and excited states of $^{48}$Ti. 
These techniques can be used 
to make predictions
for $^{76}$Ge, $^{82}$Se using the $jj44$ model space ($0f_{5/2},\ 1p,\ 0g_{9/2}$), 
and for $^{128}$Te, $^{130}$Te and $^{136}$Xe using the $jj55$ model space.

We reviewed the main contributing mechanisms to the $0 \nu \beta \beta$ decay, and we showed that
based on the present constraints from colliders one could reduce the contribution to the $0 \nu \beta \beta$
half-life to the relevant terms described in Eq. (\ref{t0red}). A reliable analysis of the
$0 \nu \beta \beta$ decay experimental data requires accurate calculations of the associated
NME.
We extended our recent analysis \cite{mh10-prc2} of the $0 \nu \beta \beta$ NME for $^{48}$Ca 
to include the heavy neutrino exchange NME, the long range gluino exchange NME, and the squark-neutrino mechanism NME.
We also presented for the first time shell model results of these new NME for the $0 \nu \beta \beta$
transitions to the g.s. and the first excited $0^+_2$ state in $^{48}$Ti. 

To extend this analysis to the $A > 48$ cases, 
more efforts have to be done to include all spin-orbit partners in the valence
space and satisfy the Ikeda sum-rule, reduce the center-of-mass spurious contributions, and better understand the changes in the
effective $0 \nu \beta \beta$ transition operators \cite{eh-09,schwenk-11}. 
In addition, the closure approximation used to calculate the NME 
within the shell model approach and by other methods (e.g. IBA-2 \cite{iba-09}, PHFB \cite{phfb}, and
GCM \cite{gcm}) needs to be further checked for accuracy, especially for the heavy neutrino
exchange, the long range gluino exchange, and the squark-neutrino mechanism. An analysis of the 
double beta decay of $^{136}$Xe that addresses some of these issues is in preparation.

\section{Appendix}

The matrix elements for  the light and heavy neutrino exchange 
in Eq. (\ref{t0red}) have the same structure as that described in Eqs. (3)-(6) of Ref.
\cite{mh10-prc2}. For $M^{0 \nu}_{\nu}$ the neutrino potential is the same as in Eq. (7)
of \cite{mh10-prc2}

\begin{eqnarray}
H_\alpha(r) & = & \frac{2R}{\pi}\int^\infty_0 f_{\alpha}(qr)
\frac{h_\alpha(q^2)}{q+\left< E \right> }G_\alpha(q^2) q dq  ,
\label{menu}
\end{eqnarray}

\noindent
with the same ingredients described in Eqs. (9)-(12) of \cite{mh10-prc2}. Here
we corrected the $(\mu_p-\mu_n)$ value to 4.71, an error that seems to be propagating
for some time through the literature \cite{rmp-08}. This correction explains the
small difference between the $M^{0 \nu}_{\nu}$ values of Table II and corresponding ones 
reported in Ref. \cite{mh10-prc2}.
Fortunately, this correction only changes the 
matrix elements by few percents. All other constants are the same as in Ref. \cite{mh10-prc2}.
In particular, we used $g_A=1.254$ and $R=1.2 A^{1/3}$ fm.
For the $M^{0 \nu}_N$ there is a slight change in the
neutrino potentials

\begin{eqnarray}
H_\alpha(r) & = & \frac{2R}{\pi m_e m_p}\int^\infty_0 f_{\alpha}(qr)
h_\alpha(q^2)G_\alpha(q^2) q^2 dq  ,
\label{meN}
\end{eqnarray}

\noindent
where $m_e$ and $m_p$ are the electron and proton mass, respectively.

The most significant contributions to $M^{0 \nu}_{\lambda'}$ and 
$M^{0 \nu}_{\tilde{q}}$ have a similar structure as $M^{0 \nu}_{\nu}$ and $M^{0 \nu}_N$,
however, only the $\alpha=GT,T$ terms in Eq. (\ref{sme}) are contributing. 
The radial neutrino
potentials for $M^{0 \nu}_{\lambda'}$ have the same form as those used for
$M^{0 \nu}_N$, Eq. (\ref{meN}), but with different $h_{\alpha}$:

\begin{equation}
h_{GT,T}=-\left(c^{1\pi}+c^{2\pi}\right) 
\left[\frac{m_em_pq^2/m_{\pi}^4}{1+q^2/m_{\pi}^2}+
\frac{2m_em_pq^2/m_{\pi}^4}{\left(1+q^2/m_{\pi}^2\right)^2} \right],
\end{equation}

\noindent
where $m_\pi$ is the charged pion mass, 139 MeV. 
Expressions for $c^{1\pi}$ and $c^{2\pi}$
are given in Ref. \cite{ves12}. The numerical values we used are
$c^{1\pi}=-85.23$ and $c^{2\pi}=368.0$.

The radial neutrino
potentials for $M^{0 \nu}_{\tilde{q}}$ have the same form as those used for
$M^{0 \nu}_{\nu}$, Eq. (\ref{menu}), but with different $h_{\alpha}$:

\begin{equation}
h_{GT,T}=-\frac{1}{6}\frac{m_{\pi}^2}{m_e\left(m_u+m_d\right)}
\frac{q^2/m_{\pi}^2}{\left(1+q^2/m_{\pi}^2\right)^2},
\end{equation}

\noindent
where $m_u$ and $m_d$ are the current up and down quark masses. In the calculation
we used $m_u+m_d=11.6$ MeV.

\begin{acknowledgments}
The  author had useful conversations with B.A. Brown and
S. Stoica. Support from the US NSF Grant PHY-1068217 and 
the SciDAC Grant NUCLEI is acknowledged.
\end{acknowledgments}

\end{document}